\documentclass[a4paper,11pt]{article}
\usepackage{pos}
\usepackage{caption}
\usepackage{subcaption}

\title{The density of state method  for first-order phase transitions in Yang-Mills theories}

\author*[a]{David Mason}
\author[b,c]{Biagio Lucini}
\author[a]{Maurizio Piai}
\author[d,e,f,g]{Enrico Rinaldi}
\author[h]{Davide Vadacchino}

\affiliation[a]{Department of Physics, Faculty of Science and Engineering, Swansea University (Park Campus),
Singleton Park, SA2 8PP Swansea, Wales, United Kingdom}
\affiliation[b]{Department of Mathematics, Faculty of Science and Engineering, Swansea University (Bay Campus),
Fabian Way, SA1 8EN Swansea, Wales, United Kingdom}
\affiliation[c]{Swansea Academy of Advanced Computing, Swansea University (Bay Campus), Fabian Way, SA1 8EN Swansea, Wales, United Kingdom}
\affiliation[d]{Physics Department, University of Michigan, Ann Arbor, MI 48109, USA}
\affiliation[e]{Theoretical Quantum Physics Laboratory, Cluster of Pioneering Research, RIKEN, Wako, Saitama 351-0198, Japan}
\affiliation[f]{Interdisciplinary Theoretical \& Mathematical Science Program, RIKEN (iTHEMS), 2-1 Hirosawa, Wako, Saitama, 351-0198, Japan}
\affiliation[g]{Center for Quantum Computing (RQC), RIKEN, Wako, Saitama 351-0198, Japan}
\affiliation[h]{Centre for Mathematical Science, University of Plymouth, Plymouth, PL4 8AA, United Kingdom}

\emailAdd{2036508@Swansea.ac.uk}
\emailAdd{b.lucini@swansea.ac.uk}
\emailAdd{m.piai@swansea.ac.uk}
\emailAdd{erinaldi.work@gmail.com}
\emailAdd{davide.vadacchino@plymouth.ac.uk}

\abstract{Lattice Field Theory can be used to study finite temperature first-order phase transitions in new, strongly-coupled gauge theories of phenomenological interest. Metastable dynamics arising in proximity of the phase transition can lead to
large, uncontrolled numerical errors when analysed  with standard methods.
In this contribution, we discuss a prototype lattice calculation in which the first-order deconfinement transition in the strong Yang-Mills sector of the standard model is analysed using a novel lattice method, the logarithmic linear relaxation algorithm. This method provides a determination of the density of states of the system with exponential error suppression.}

\FullConference{%
The 39th International Symposium on Lattice Field Theory,\\
8th-13th August, 2022,\\
Rheinische Friedrich-Wilhelms-Universität Bonn, Bonn, Germany
}

\begin{document}

\newcommand{\sun}{$SU(N_c)$~}
\newcommand{\suthree}{$SU(3)$~}

\makeatletter
\newsavebox{\@brx}
\newcommand{\llangle}[1][]{\savebox{\@brx}{\(\m@th{#1\langle}\)}%
  \mathopen{\copy\@brx\mkern2mu\kern-0.9\wd\@brx\usebox{\@brx}}}
\newcommand{\rrangle}[1][]{\savebox{\@brx}{\(\m@th{#1\rangle}\)}%
  \mathclose{\copy\@brx\mkern2mu\kern-0.9\wd\@brx\usebox{\@brx}}}
\makeatother

\maketitle
\section{Introduction}
\label{intro}
\sun gauge theories at finite temperatures are known to undergo a deconfinement phase transition~\cite{mclerran1981quark}. Non-perturbative studies of this transition give valuable insights on the dynamics of Yang-Mills theories from a number of perspectives. For instance, one can characterise the behaviour of thermodynamic observables as a function of the number of colours $N_c$~\cite{Lucini:2003zr,lucini2005properties,Panero:2009tv}. First-order phase transitions in the early universe leave an imprint in gravitational waves (see e.g., Refs.~\cite{huang2021testing,Halverson:2020xpg,Kang:2021epo,Reichert:2022naa}). This opens the exciting possibility of using gravitational waves as additional probes of physics beyond the standard model. Among other applications, this programme is relevant for extensions of the standard model that propose a composite nature for the Higgs field, new top-quark partners, or dark matter candidates, such as those based on $Sp(4)$ gauge theories, recently studied numerically in, e.g., Refs.~\cite{Bennett:2017kga,Bennett:2019jzz,Bennett:2019cxd,Bennett:2022yfa,Kulkarni:2022bvh}. To understand the strength of the gravitational waves originated by the phase transition in a given theory, non-perturbative calculations of relevant observables need to be performed. In this contribution, we report on a calculation using the the linear logarithmic relaxation (LLR) algorithm~\cite{langfeld2012density} in \suthree Yang-Mills. For this system, a high-precision calculation of the latent heat has been recently provided in Ref.~\cite{Borsanyi:2022xml} . A calculation using a similar methodology to the one we discuss here but targeting $SU(4)$ has been discussed in Ref.~\cite{springer2021density}. Parts of this work have been reported already in Ref.~\cite{Mason:2022trc}, to which we refer the reader for complementary discussions. A more extended publication is in preparation~\cite{inpreparation}.

The rest of this work is structured as follows. In Sect.~\ref{sec:meta} we provide a description of the lattice system, an exposition of the algorithm and a discussion of the numerical implementation. Section~\ref{sec:results} reports on our numerical findings. Finally, our conclusions and an overview of future work are given in Sect.~\ref{sec:conclusions}.

\section{Lattice setup and LLR simulation details}
\label{sec:meta}
We consider a system discretised on an asymmetric lattice of size $\tilde{V}/a^4 = N_t \times N_s^3$, with $a$ the lattice spacing. For convenience, we set $a = 1$. The degrees of freedom are \sun gauge fields defined on the links of the lattice, $U_{\mu}(i)$. The model is described by the path integral
\begin{equation}
    \label{eqn:lattice_action}
Z(\beta) = \int {\cal D} U_\mu(i) e^ {- \beta S} \ ,  \ S =  \sum_{j=0}^{\tilde{V}} \sum_{\mu; \nu > \mu}\left(1 - \frac{1}{N_c} \Re(\textrm{Tr}[U_{\mu\nu}(j)])\right) \,,
\end{equation}
where $S$ is the Wilson action, with the sum running over the real component of the trace of all the plaquettes, $U_{\mu\nu}(j)$, and $\beta = 2 N_c/g_0^2$, with $g_0$ the bare lattice gauge coupling. The finite temperature setup is given by the condition $N_s \gg N_t$, and the temperature $T$ is set by $N_t$ and  $a$, as $T=(N_t a)^{-1}$. 

\sun gauge theories undergo a deconfinement phase transition at some critical value of the temperature $T_c$ (or, equivalently, of the coupling $\beta_c$). An order parameter for the transition is the Polyakov loop vacuum expectation value, $\langle l_p \rangle$, which detects the breaking of the $\mathbb{Z}_{N_c}$ center symmetry. For $N_c \ge 3$ the deconfinement phase transition is first order. A general feature of first-order phase transitions at criticality is the coexistence of phases. Because of the free energy barrier between the two equilibrium states, widely used local Monte Carlo update methods such as the Metropolis and the heat-bath algorithms, have correlation times that grow exponentially with the volume. To overcome this issue, we explore the use of the LLR method~\cite{langfeld2012density}, which has been demonstrated to efficiently sample systems near criticality at first-order phase transitions~\cite{langfeld2016efficient,lucini2016overcoming}. 

The LLR method is based on the determination of a suitable approximation of the density of states $\rho(E)$ as a function of the energy $E$ through samplings restricted to energy intervals of fixed widths $\delta_E$ in a dynamically relevant energy range $[E_{min},E_{max}]$. In particular, the approximation is provided in terms of a continuous piecewise function in each of the $N+1$ intervals centered at energy values $E_n$,  
\begin{equation}
    \label{eqn:RM_rho}
    \rho (E) \approx  \rho_0 \exp\left(\sum_{k=0}^{n-1}(a_k \delta_E) + a_n (E-E_n + \delta_E / 2)\right) \ , \qquad E_n - \delta_E/2 \le E \le E_n + \delta_E/2 \ .
\end{equation}
Here $\rho_0$ is a normalisation constant that drops out when computing averages of thermodynamic observables and can hence be fixed arbitrarily.
In order to compute the $a_n$, which are the a priori unknown quantities appearing in the above relationship, expectation values of observables $O$ restricted to the interval $[E_n - \delta_E/2, E_n + \delta_E/2]$ are defined as
\begin{align}
\llangle O(E) \rrangle_{a_n} &= {\cal N}^{-1} 
\int [ D \phi ] O(\phi) e^{-a_n S [ \phi ]} \left( \theta(S[\phi] - E_n + \delta_E/2) -  \theta(S[\phi] -  E_n - \delta_E/2) \right) \ , \\ 
{\cal N} &= \int [ D \phi ] e^{-a_n S [ \phi ]} \left( \theta(S[\phi] - E_n + \delta_E/2) -  \theta(S[\phi] - E_n - \delta_E/2) \right) \ . 
\end{align}
We refer to those energy restricted variables as double angle expectation values. With these definitions, $a_n$ is the solution of the stochastic equation
\begin{equation}
    \llangle \Delta E \rrangle_{a_n} =\llangle E - E_n \rrangle_{a_n} = 0 \ , 
\end{equation}
which is determined with the Robbins-Monro~\cite{RobbinsMonro:1951} iterations
\begin{equation}
    \label{RM_ak}
    \llangle \Delta E \rrangle_{a_n} =\llangle E - E_n \rrangle_{a_n} = 0
    , \qquad a_n^{(m+1)} = a_n^{(m)} - \frac{12 \llangle \Delta E \rrangle_{a_n^{(m)}}}{\delta_E^2 (m+1)} \ . 
\end{equation}
$a_n^{(m)} \to a_n$ in the limit $m \to \infty$. Note that $a_n = 1/t_n$, with $t_n$ the micro-canonical inverse temperature associated with the energy value $E_n$. In the numerical determination of the $a_n$, the systematics related to the truncation of the Robbins-Monro (RM) recursion is handled by repeating the calculation for each $a_n$ at fixed value of number of steps $m$ and bootstrapping the result in any subsequent analysis. This enables us to swap the unknown truncation systematics with an easier to treat statistical error. Ergodicity in our calculation is recovered via umbrella sampling, as described in Ref.~\cite{lucini2016overcoming}, according to which the size of the intervals are increased to $\Delta_E = 2\delta_E$, and consecutive intervals are given an overlap region, $ (E_{n+1} - E_n)$. If two lattices in adjacent intervals are both in the overlap region after a Robbins-Monro iteration, a swap of configurations between the two intervals is attempted with a Metropolis step. When using umbrella sampling, the factor of $\delta_E$ in equation \ref{RM_ak} should be replaced with $\Delta_E$. 

In this contribution we report on a numerical study for $N_c = 3$ on a $N_t \times N_s^3 = 4 \times 20^3$ lattice. At this size, the system is tractable also with conventional Monte Carlo simulations consisting of an admixture of heath-bath and Metropolis steps. This enables us to test the methodology comparing relevant LLR results with more standard calculations. As the energy $E$, we take the value of the action on a given configuration. The energy boundaries have been chosen to be $E_{min}/6\tilde{V} = 0.439487341$ and $E_{max}/6\tilde{V} = 0.459698522$, with the energy interval divided into 55 subintervals of width $\Delta_E/6\tilde{V} = 0.000748562$. 

\begin{figure}
\includegraphics[width=0.5\textwidth]{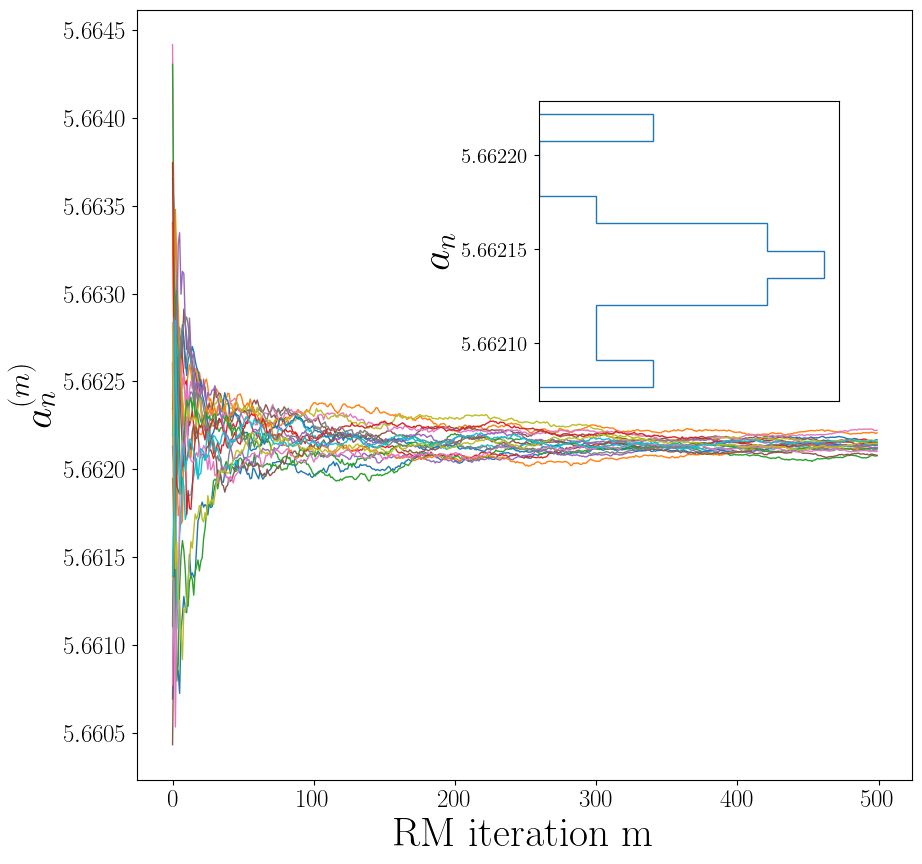}
\centering
\caption{A sample of 20 RM sequences at a fixed value of $E_n$. Different colours are used for different trajectories. The inset shows the obtained $a_n^{(m)}$ distribution at the truncation value $m = 500$.}
\label{fig:RM_many}       
\end{figure}

\begin{figure}
\includegraphics[width=0.5\textwidth]{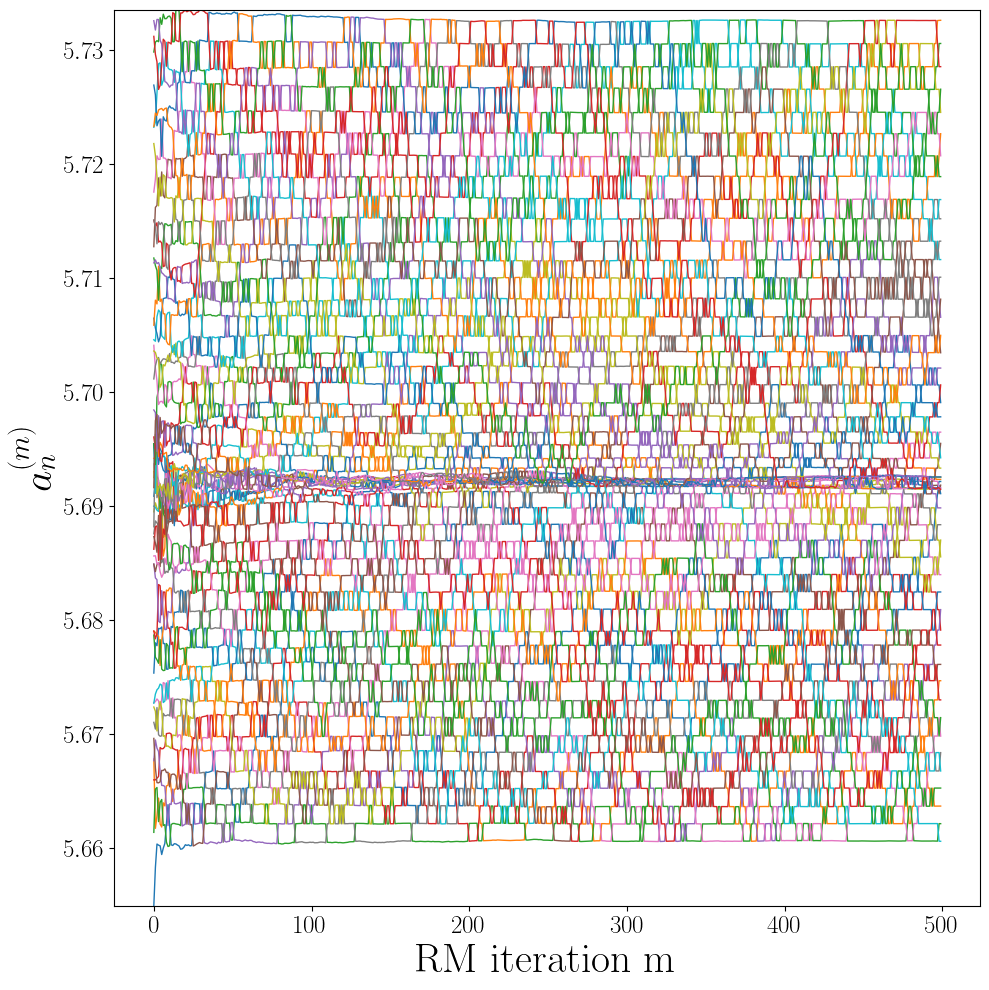}
\centering
\caption{Trajectories for all intervals on a single repeat of the Robbins-Monro algorithm that includes the replica swapping. The $x$ axis shows the RM iterations $m$, while the $y$ axis displays the value $a_n^{(m)}$. The colour follows the history of a lattice as it is swapped between overlapping intervals.}
\label{fig:RM_umbrella}       
\end{figure}

An example of 20 repeats of a Robbins-Monro's trajectory is shown in Fig.~\ref{fig:RM_many}. The plot provides good evidence that for sufficiently large numbers of iterations $m$ the sequences are normally distributed around the asymptotic value, with a variance that decreases with $m$, in agreement with the arguments reported in Ref.~\cite{langfeld2016efficient}. Fig.~\ref{fig:RM_umbrella} shows the effectiveness of the replica exchange. While in general the algorithm is efficient at swapping configurations across intervals, we note that trajectories become dense at criticality. This is due to the dynamics of first-order phase transitions, which has been investigated in Ref.~\cite{neuhaus20032d}.

\section{Thermodynamic observables}
\label{sec:results}
Using the density of states calculated with the LLR algorithm, the canonical information on the system can be recovered. If the observable $O$ depends on the energy, we can write the canonical expectation value as the ratio of two numerical integrals, 
\begin{equation}\label{eq:vev_obs_2}
  \langle O(E) \rangle _\beta= \frac{1}{Z(\beta)}
 \int \rho(E) O(E)e^{-\beta E}  \ , \qquad  Z(\beta) = \int \rho(E) e^{-\beta E} \ , 
\end{equation}
with $\rho(E)$ approximated using the expression in Eq.~(\ref{eqn:RM_rho}), where the limit $\delta_E \to 0$ should be taken. Here we work at finite $\Delta_E/6\tilde{V} = 0.000748562$, having checked that for this value corrections in $\delta_E$ are negligible with respect to the quoted statistical errors. The integrals appearing in Eq.~(\ref{eq:vev_obs_2}) should extend over the whole range of allowed  energies. However, standard thermodynamic arguments state that only states around the energy of interest contribute. The range  $E/6\tilde{V} \in [0.439487341, 0.459698522]$ has been chosen so that for all the $\beta$ values of interest the contributions close to the boundaries are negligible. 

\begin{figure}
     \centering
     \begin{subfigure}[b]{0.49\textwidth}
         \centering
         \includegraphics[width=\textwidth]{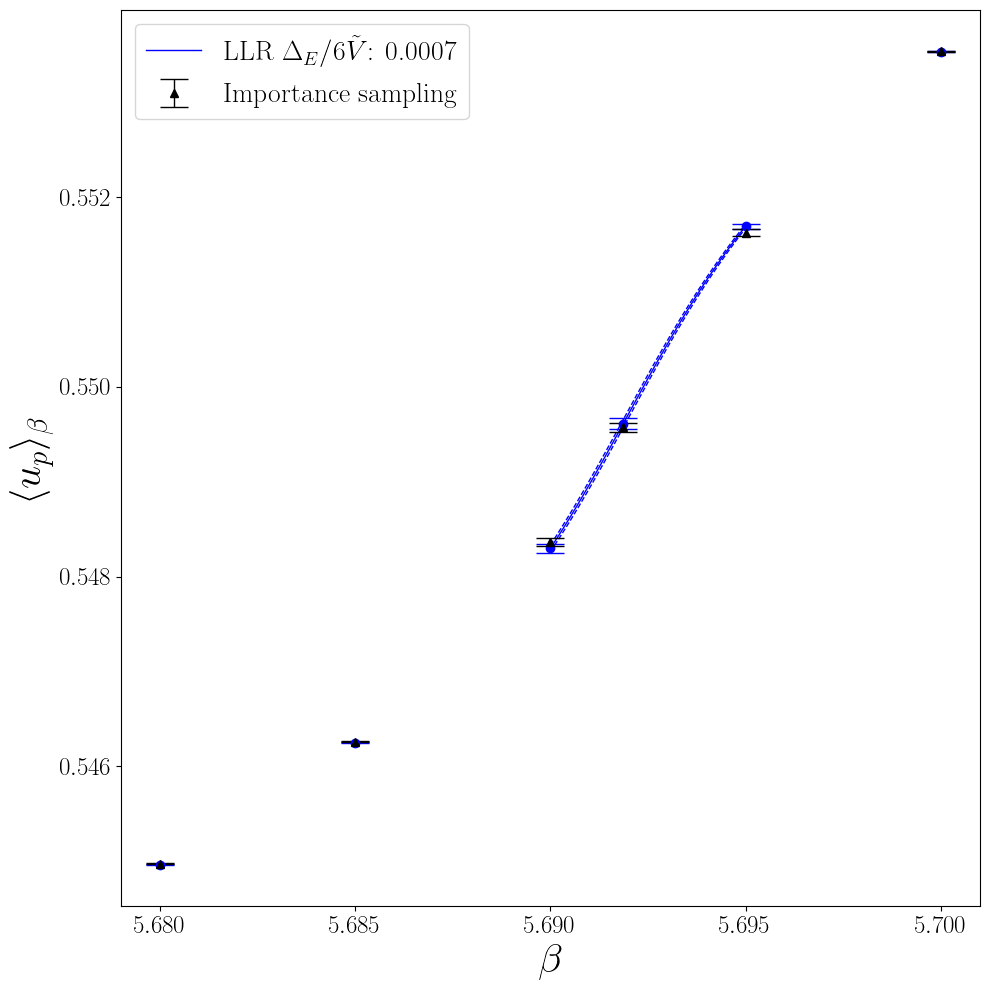}
         \caption{}
         \label{fig:plaq}
     \end{subfigure}
     \hfill
     \begin{subfigure}[b]{0.49\textwidth}
         \centering
         \includegraphics[width=\textwidth]{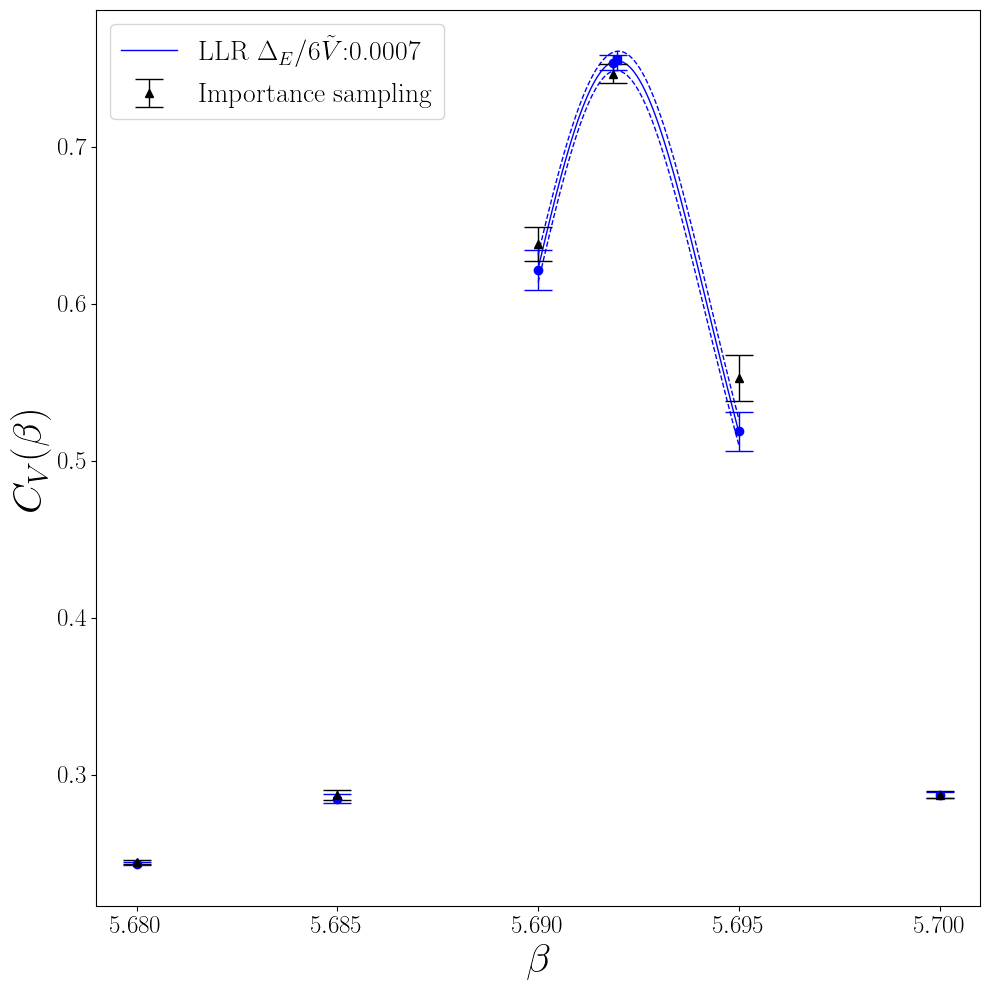}
         \caption{}
         \label{fig:Cv}
     \end{subfigure}
        \caption{Thermodynamic observables measured with the LLR method (blue circles), compared to results from standard importance sampling (black triangles), for \suthree Yang-Mills gauge theory on a $4 \times 20^3$ lattice. The blue curves are reconstructed observables from the LLR method with a finer resolution in $\beta$, restricted to the region around the phase transition. Left panel:  average plaquette $\langle u_p \rangle_\beta$ against the coupling $\beta$. Right panel: specific heat $C_V(\beta) \equiv \langle u_p^2 \rangle_\beta - \langle u_p \rangle_\beta^2$ against the coupling $\beta$.}
        \label{fig:Full_plaquette}
\end{figure}

Two observables that depend on $E$ and are widely studied to characterise the phase transition are the plaquette expectation
\begin{equation}
\langle u_p \rangle_\beta = 1 - \langle E \rangle _\beta /6 \tilde{V}
\end{equation}
and the specific heat
\begin{equation}
  C_V(\beta) \equiv \langle u_p^2 \rangle_\beta - \langle u_p \rangle^2_\beta \ , 
\end{equation}
the latter being the fluctuations of the former. Their reconstruction near criticality using the LLR determined density of states is shown in Fig.~\ref{fig:Full_plaquette}. These values are compared to results from a computation using standard lattice methods with 500,000 configurations. The plots show good agreement between the two determinations, with the LLR having the advantage of providing a dense set of points at negligible additional cost, since, once the approximated $\rho(E)$ has been determined, $\beta$ is just a parameter in the numerical integration. 

\begin{figure}
     \centering
     \begin{subfigure}[b]{0.49\textwidth}
         \centering
         \includegraphics[width=\textwidth]{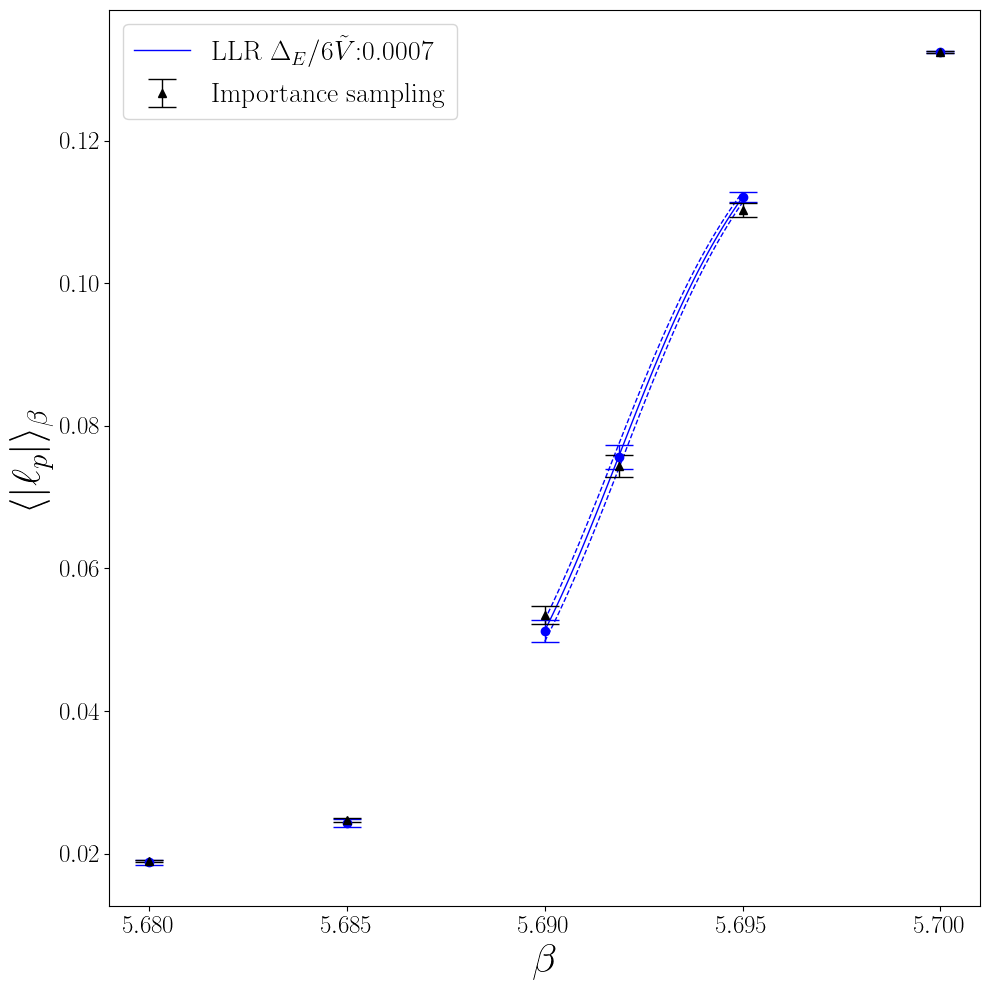}
         \label{fig:lp}
     \end{subfigure}
     \hfill
     \begin{subfigure}[b]{0.49\textwidth}
         \centering
         \includegraphics[width=\textwidth]{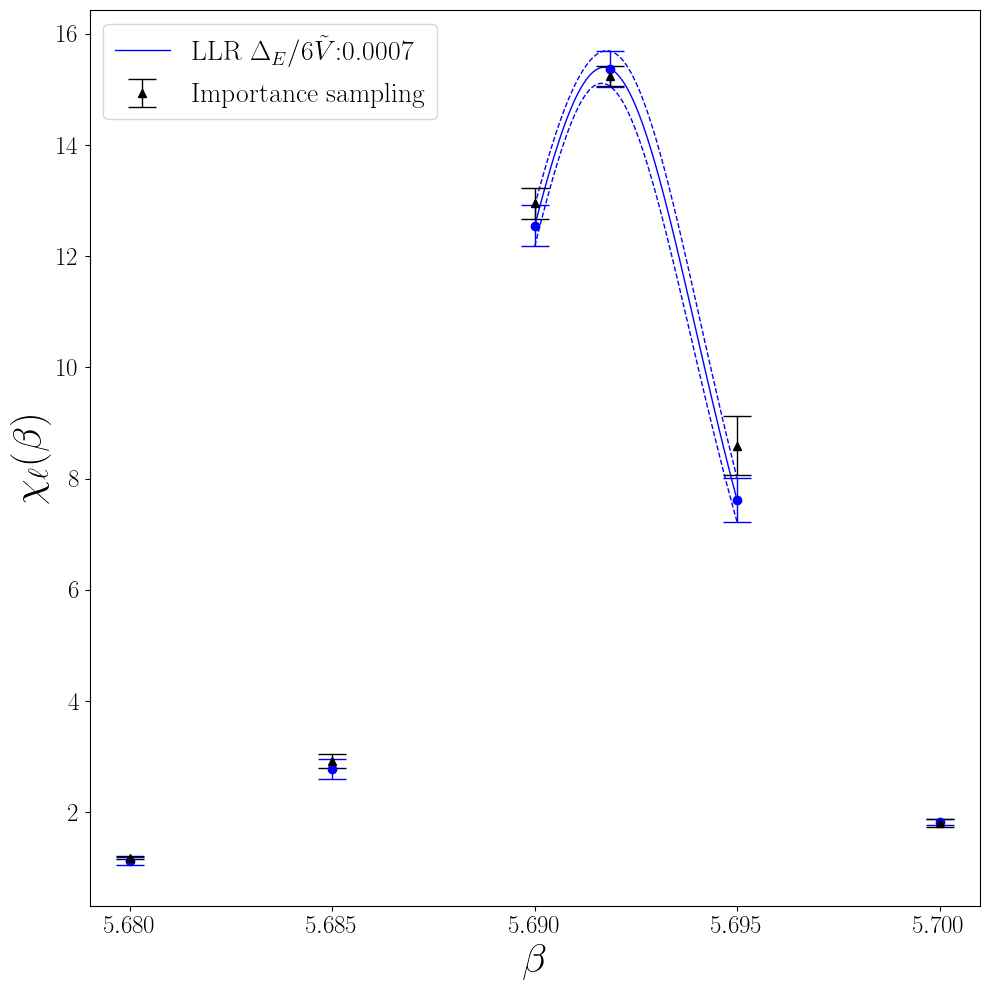}
         \label{fig:Xl}
     \end{subfigure}
        \caption{Thermodynamic observables measured with the LLR method (blue circles), compared to results from standard importance sampling (black triangles), for \suthree Yang-Mills gauge theory on a $4 \times 20^3$ lattice. The blue curves are reconstructed observables from the LLR method with a finer resolution in $\beta$, restricted to the region around the phase transition. Left panel:  average absolute value of the Polyakov loop $\langle |l_p| \rangle_\beta$ against the coupling $\beta$. Right panel: Polyakov loop susceptibility $\chi_l (\beta)\equiv N_s^3(\langle |l_p|^2 \rangle_\beta - \langle |l_p| \rangle_\beta^2)$ against the coupling $\beta$.}
        \label{fig:Full_polyakov}
\end{figure}

Following Refs.~\cite{langfeld2016efficient,Cossu:2021bgn}, we can also compute general observables $B$ that do not have an explicit functional form as a function of $E$ at a given coupling $\beta$ by carrying out measurements of $B[\phi]$ on configurations sampled at an inverse temperature $a_n$, with the sampling restricted to the interval centered at $E_n$. The reconstructed canonical expectation of the observable is then 
\begin{equation}\label{eq:vev_gen}
 \langle B[\phi] \rangle_\beta = \frac{1}{Z(\beta)} \sum_{n=0}^{N} \delta_E \rho(E_n) \llangle B[\phi]\exp{\left( -\beta S[\phi] + a_n (S[\phi] - E_n)\right)}\rrangle_{a_n}.
\end{equation}
Once the full set of $a_n$ values were found, we generated 2000 configurations and measured the action $S \equiv E$ and the absolute value of the average Polyakov loop $\langle |\ell_p| \rangle _\beta$. The average Polyakov loop and the Polyakov loop susceptibility were determined with the LLR reconstruction prescription and compared against standard importance sampling methods, as shown in Fig. \ref{fig:Full_polyakov}. Once again, we found excellent agreement between the two methods,

\begin{figure*}
\centering
\includegraphics[width=0.5\textwidth]{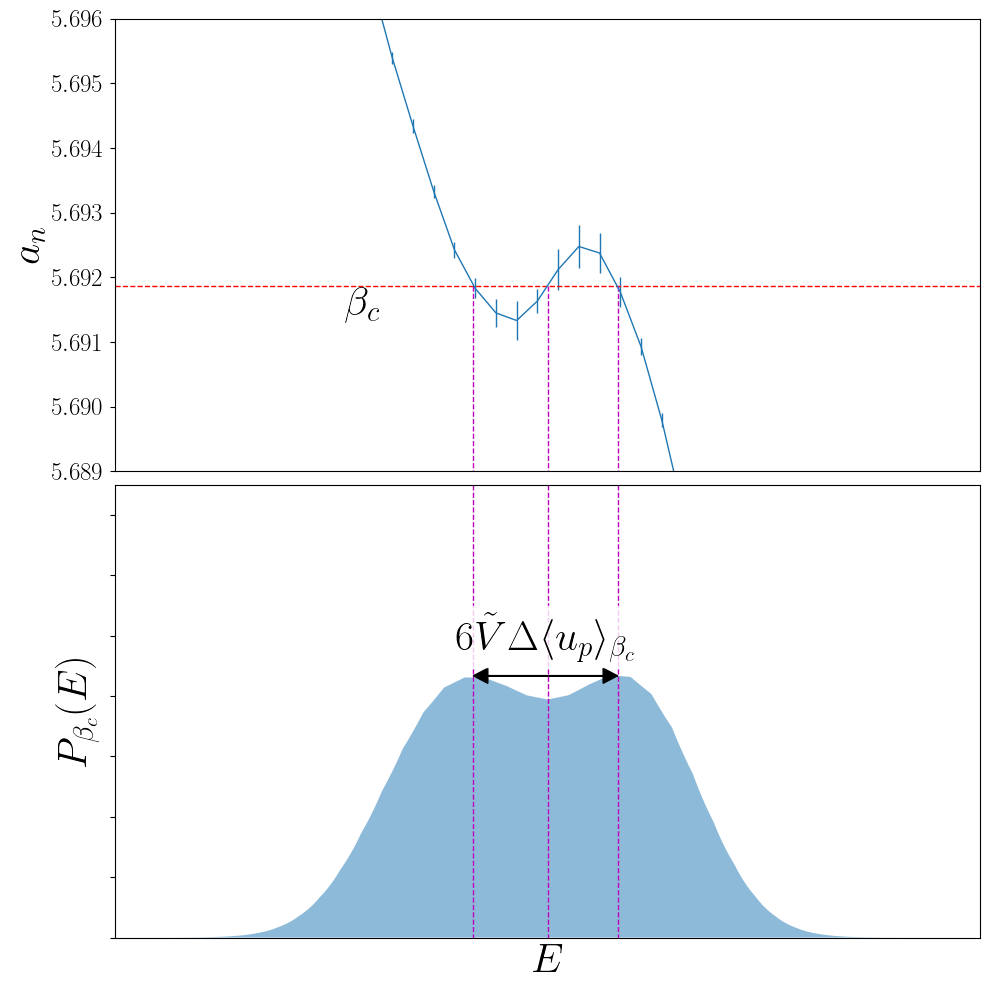}
\caption{Results of the LLR analysis of \suthree Yang-Mills gauge theory on a $4 \times 20^3$ lattice at the values of the LLR parameters given in the text. Top panel: values of the $a_n$ against the centres of the energy intervals $E_n$, with a linear interpolation between the points.
Bottom panel: reconstructed probability distribution $P_{\beta_c}(E)$ of the energy $E$ at the critical coupling $\beta_c$.
The horizontal dashed line shows the value of the critical coupling, and the vertical lines are the average energy values at which $a_n = \beta_c$, which correspond to the locations of the extrema of the probability distribution.}
\label{fig:energy_distribution}       
\end{figure*}

Investigating the lattice system using the LLR method gives us access to its microcanonical information. From this the probability distribution of the system at coupling $\beta$, $P_\beta (E)$ can be determined through the equations
\begin{equation}
    \label{Pot}
     P_\beta(E) = \frac{1}{Z}\rho(E)e^{-\beta E} \ .
\end{equation}
This quantity is displayed in Fig.~\ref{fig:energy_distribution}, together with the values of $a_n$ determined in the relevant energy interval, at the value of $\beta$ for which the two peaks have been found to have equal height, which we take as a definition of $\beta_c$. The distance between the two peaks determines the strength of the transition through the latent heat. 

The density of states can be linked to the entropy $s$ of the lattice system, $s = \log \rho(E)$. With this definition, and remembering that $a_n = 1/t_n$, the free energy of the thermodynamic system is then obtained as
\begin{equation}
    \label{therm}
    F =E - ts \ .
\end{equation}

\begin{figure}
\includegraphics[width=0.5\textwidth]{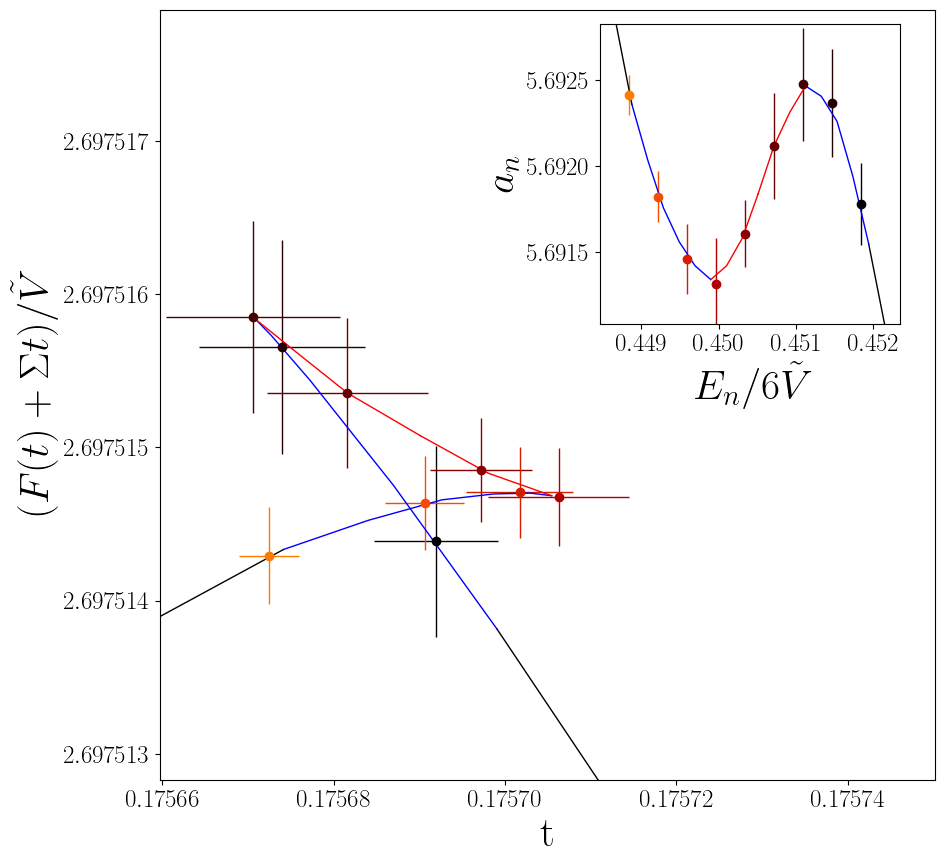}
\centering
\caption{The free-energy for the \suthree Yang-Mills gauge theory on a $4 \times 20^3$
  lattice, computed with the LLR method. $\Sigma$ is a constant computed as the temperature average of $s  - \ln \rho_0$, where $\rho_0$, as discussed in Ref.~\cite{Mason:2022trc}.}

  \label{fig:free_energy}       
\end{figure}

The behaviour of free-energy around the critical point shows the expected swallow-tail structure that indicates the meta-stable first-order behaviour (see., e.g., Ref.~\cite{Kubiznak:2012wp}), as represented in Fig.~\ref{fig:free_energy}. The calculated values are obtained through the subtraction of a linear term to remove the effect of the choice of $\rho_0$ in the density of states, as shown in Ref.~\cite{Mason:2022trc}. The critical point, i.e., the point at which both phases are equally likely, corresponds to the point at which the free-energy curve intersects itself. 

\section{Conclusion and outlook}
\label{sec:conclusions}
Motivated by the increasing interest in gravitational-wave signatures of early universe, first-order phase transitions, we have begun the investigations of thermal Yang-Mills theories using the LLR algorithm.
This method avoids long correlations due to metastabilities near criticality, and hence produces more robust results with a contained calculation cost. Studying \suthree gauge theory, we have benchmarked the LLR calculation on a lattice for which local Monte Carlo updates have a bearable cost, finding agreement between this method and more conventional approaches. At the same time, we have provided accurate results for the probability distribution of the energy at fine resolution and for the free energy, which is not accessible at our calculated precision with traditional methods. 

\acknowledgments
We would like to thank David Schaich and Felix Springer for discussions.
The work of D.~V. is partly supported by the Simons Foundation under the program ``Targeted Grants to Institutes'' awarded to the Hamilton Mathematics Institute.
The work of D.~M. is supported by a studentship awarded by the Data Intensive Centre for Doctoral Training, which is funded by the STFC grant ST/P006779/1.  
E.~R. was supported by Nippon Telegraph and Telephone Corporation (NTT) Research.
The work of B.~L. and M.~P. has been supported in part by the STFC 
 Consolidated Grants No. ST/P00055X/1 and No. ST/T000813/1. B.L. and M.P. received funding from the European Research Council (ERC) under the European Union’s Horizon 2020 research and innovation program under Grant Agreement No.~813942. The work of BL is further supported in part by the Royal Society Wolfson Research Merit Award WM170010 and by the Leverhulme Trust Research Fellowship No. RF-2020-4619. Numerical simulations have been performed on the Swansea SUNBIRD cluster (part of the Supercomputing Wales project) and AccelerateAI A100 GPU system, and on the DiRAC Data Intensive service at Leicester. The Swansea SUNBIRD system and AccelerateAI are part funded by the European Regional Development Fund (ERDF) via Welsh Government. The DiRAC Data Intensive service at Leicester is operated by the University of Leicester IT Services, which forms part of the STFC DiRAC HPC Facility (www.dirac.ac.uk). The DiRAC Data Intensive service equipment at Leicester was funded by BEIS capital funding via STFC capital grants ST/K000373/1 and ST/R002363/1 and STFC DiRAC Operations grant ST/R001014/1. DiRAC is part of the National e-Infrastructure.\\
{\bf Open Access Statement - } For the purpose of open access, the authors have applied a Creative Commons 
Attribution (CC BY) licence  to any Author Accepted Manuscript version arising.

\bibliographystyle{jhep}
\bibliography{references}
\end{document}